\documentclass{elsart}
\usepackage{graphicx}
\usepackage{amsmath}
\usepackage{amssymb}
\usepackage{epsfig}
\usepackage{dcolumn}% Align table columns on decimal point
\usepackage{bm}% bold math
\begin{document}
\begin{frontmatter}

\title{Multifractal spectrum of phase space \break related to generalized thermostatistics}

\author{A.I. Olemskoi}
\ead{alex@ufn.ru}
\address{Institute of Applied Physics, Nat. Acad. Sci. of Ukraine \break 58, Petropavlovskaya St., 40030 Sumy,
Ukraine}
\author{V.O. Kharchenko}
\ead{vasiliy@imag.kiev.ua}
\address{Institute of Magnetism, Nat. Acad. Sci. of Ukraine \break
36-B, Vernadsky St., 03680 Kyiv, Ukraine}
\author{V.N. Borisyuk}
\ead{vadimborisyuk@yahoo.com}
\address{Sumy State University \break 2, Rimskii-Korsakov St., 40007 Sumy, Ukraine}

\date{}

\begin{abstract}
We consider a self-similar phase space with specific fractal dimension $d$
being distributed with spectrum function $f(d)$. Related thermostatistics is
shown to be governed by the Tsallis formalism of the non-extensive statistics,
where the non-additivity parameter is equal to ${\bar\tau}(q)\equiv
1/\tau(q)>1$, and the multifractal function $\tau(q)= qd_q-f(d_q)$ is the
specific heat determined with multifractal parameter $q\in
\left[1,\infty\right)$. In this way, the equipartition law is shown to take
place. Optimization of the multifractal spectrum function $f(d)$ derives the
relation between the statistical weight and the system complexity. It is shown
the statistical weight exponent $\tau(q)$ can be modeled by hyperbolic tangent
deformed in accordance with both Tsallis and Kaniadakis exponentials to
describe arbitrary multifractal phase space explicitly. The spectrum function
$f(d)$ is proved to increase monotonically from minimum value $f=-1$ at $d=0$
to maximum one $f=1$ at $d=1$. At the same time, the number of monofractals
increases with growth of the phase space volume at small dimensions $d$ and
falls down in the limit $d\to 1$.
\end{abstract}

\end{frontmatter}

\section{Introduction}\label{sec:level1}

A generalization of the statistical mechanics onto the non-extensive
thermostatistics is known to be based on the deformation procedure of both
logarithm and exponential functions \cite{Tsallis,Naudts,1}. The simplest way
to introduce these functions into the thermostatistics scheme is to consider
the equation of motion for dimensionless volume
$\gamma=\Gamma/(2\pi\hbar)^{6N}$ of the supported phase space ($\hbar$, $N$
being Dirac-Planck constant and particle number). In the course of evolution of
the ensemble with statistical weight $w=w(\gamma)$ and entropy $S$, the

variation rate of the phase space volume is governed by the equation \cite{1}
\begin{equation}
\frac{{\rm d}\gamma}{{\rm d}t}=w(\gamma)\frac{{\rm d}S}{{\rm d}t}.\label{1}
\end{equation}
Following from here relation ${\rm d}S={\rm d}\gamma/w(\gamma)$ gives the
entropy corresponding to the whole statistical weight $W$:
\begin{equation}
S(W)=\int\limits_{\gamma(1)}^{\gamma(W)}\frac{{\rm
d}\gamma}{w(\gamma)}.\label{2}
\end{equation}
Here, we take into account that entropy %$S(1)=0$
of a single state $W=1$ vanishes, i.e., $S(1)=0$.

In the case of the smooth phase space, one has trivial relation
$w(\gamma)=\gamma$ whose insertion into Eq.(\ref{2}) arrives at the Boltzmann
entropy $S=\ln W$. However, complex systems have fractal phase space with the
dimension $D<6N$, so that relation between the statistical weight and the
corresponding volume should be generalized by the power-law dependence
\begin{equation}
w(\gamma)=\gamma^d \label{3}
\end{equation}
where the specific fractal dimension $d\equiv{D}/{6N}\leq 1$ is introduced as
the exponent. Insertion of Eq.(\ref{3}) into the integral (\ref{2}) gives the
expression\footnote{This expression is equivalent to Eq.(16) in Ref.\cite{1}
since W in our manuscript denotes $\mathcal{N}$ given by (11) in \cite{1}.}
\begin{equation}
S(W)={\bar d}\ln_{2-{\bar
d}}(W),\quad\ln_q(x)\equiv\frac{x^{1-q}-1}{1-q}\label{4}
\end{equation}
which is reduced to the Tsallis logarithm $\ln_q(x)$ where the non-additivity
parameter $q$ is replaced by the difference $2-{\bar d}$ with ${\bar d}\equiv
1/d\geq 1$ being the inverse value of the specific fractal dimension $d$ of the
phase space. Naturally, this expression gives the Boltzmann entropy in the
limit $d\to 1$.

Above formalism is based on the proposition that the phase space is related to
a monofractal set determined by single dimension $d$. However, the
considerations \cite{Lyra,Jizba,Touchette} show that a complex system behaviour
can be determined by the phase space geometry, being much more complicated, in
particular multifractal. In this connection, we aim to generalize the Tsallis
thermostatistics onto the multifractal phase space with a spectrum $f(d)$. Such
a generalization for arbitrary distribution $f(d)$ is carried out in Section
\ref{sec:level2}. Related discussion shows that physical representation of the
thermostatistics based on the multifractal phase space demands of the passage
from input distribution to escort one. An optimization procedure of the
spectrum $f(d)$ is considered in Section \ref{sec:level3} to derive the
relation between the statistical weight and the system complexity. In Section
\ref{sec:level4} we show that the monotonically increasing mass exponent
$\tau(q)$, being free energy of the multifractal set \cite{1a}, is presented by
the hyperbolic tangent deformed in accordance with both Tsallis and Kaniadakis
procedures, which allow for to describe explicitly arbitrary multifractal phase
space. Section \ref{sec:level5} is devoted to consideration of the multifractal
spectrum $f(d)$ which determines the number of monofractals within the
multifractal with the specific dimension $d$. Section \ref{sec:level6}
concludes our consideration.

\section{Thermostatistics of multifractal phase space}\label{sec:level2}

According to the self-similarity condition the specific statistical weight of
the system under consideration is given by the power law function \cite{Feder}
\begin{equation}
\varpi_q(\gamma)=\gamma^{qd} \label{1a}
\end{equation}
where $q$ is the multifractal exponent, $d\equiv{D}/{6N}\leq 1$ is the specific
fractal dimension. This function should be multiplied by the number of
monofractals with dimension $d$
\begin{equation}
\mathcal{N}_d(\gamma)=\gamma^{-f(d)} \label{2a}
\end{equation}
which are contained in the multifractal whose spectrum is determined by a
function $f(d)$. As a result, whole statistical weight, being the multifractal
measure, takes the form
\begin{equation}
w_q(\gamma)\equiv\int\limits^1_0\varpi_q(\gamma)\mathcal{N}_d(\gamma)\rho(d){\rm
d}d=\int\limits^1_0\gamma^{qd-f(d)}\rho(d){\rm d}d \label{3a}
\end{equation}
where $\rho(d)$ is a density distribution over dimensions $d$. Using the method
of the steepest descent, we arrive at the power law
\begin{equation}
w_q(\gamma)\simeq\gamma^{\tau(q)}\label{4a}
\end{equation}
which generalizes the simplest relation (\ref{3}) due to replacement of the
bare fractal dimension $d$ by the multifractal function
\begin{equation}
\tau(q)= qd_q-f(d_q),\label{5a}
\end{equation}
being the mass exponent \cite{Feder}. Here, the specific fractal dimension
$d_q$ relates to given parameter $q$ to be defined by the following conditions
of the steepest descent method:
\begin{equation}
\left.\frac{{\rm d}f}{{\rm d}d}\right|_{d=d_q}=q,\quad\left.\frac{{\rm d}^2
f}{{\rm d}d^2}\right|_{d=d_q}<0. \label{6a}
\end{equation}

Above consideration shows that the passage from the monofractal phase space to
the multifractal one is obtained by replacement of the single dimension $d$ by
the monotonically increasing function $\tau(q)$, such as $\tau(0)=-1$ and
$\tau(1)=0$. Limit behaviour of the function $\tau(q)$ is characterized by the
asymptotics \cite{Feder}
\begin{equation}
\tau\propto(q-1)\ \ \text{at}\ \ 0\leq q-1\ll 1,\qquad\tau\simeq 1\ \
\text{at}\ \ q\to\infty. \label{11}
\end{equation}
A physical domain of the $q$ parameter variation is bounded by the condition
$q\geq 1$ which ensures positive values of the function $0\leq\tau(q)\leq 1$ to
guarantee growth of the specific statistical weight (\ref{4a}) with increasing
the phase space volume.

According to the entropy expression (\ref{4}) we can use well-known Tsallis
formalism of the non-extensive statistical physics where the difference
$2-{\bar\tau}(q)$ with ${\bar\tau}(q)\equiv 1/\tau(q)>1$ plays a role of the
non-additivity parameter. Thus, the entropy in dependence of the probability
distribution $ P_i$ has the form \cite{Tsallis}
\begin{equation}
S_q=-\sum\limits_{i=1}^{W_q} P_i\frac{P_i^{{\bar\tau}-1}-1}{{\bar\tau}-1}
\label{12}
\end{equation}
where the statistical weight $W_q$ is related to given value $q$ of the
multifractal exponent. With accounting the normalization conditions and the
definition of the internal energy $E_q$
\begin{equation}
\sum\limits_{i=1}^{W_q} P_i=1,\quad E_q=\sum\limits_{i=1}^{W_q}\varepsilon_i
P_i^{{\bar\tau}(q)},\label{13}
\end{equation}
the expression (\ref{12}) arrives at the generalized distribution over energy
levels $\varepsilon_i$ as follows:
\begin{equation}
 P_i=Z_q^{-1}\exp_{{\bar\tau}(q)}\left(-\beta\varepsilon_i\right),\quad
 Z_q\equiv\sum_{i=1}^{W_q}\exp_{{\bar\tau}(q)}\left(-\beta\varepsilon_i\right).
\label{14}
\end{equation}
Here, $Z_q$ is the partition function, $\beta$ is Lagrange multiplier, not
being the physical temperature, and the deformed exponential function is
determined by the expression
\begin{eqnarray}
\exp_{\bar\tau}(x)\equiv\left\{
\begin{array}{ll}
\left[1+({\bar\tau}-1)x\right]^{\frac{1}{{\bar\tau}-1}}\ {\rm at}\
1+({\bar\tau}-1)x>0,\\ 0\ \ \qquad\quad\qquad\qquad\quad\qquad\quad{\rm
otherwise}.
\end{array} \right.
\label{D}
\end{eqnarray}
Parameter q characterizes here the multifractal spectrum through the function
$\tau(q)$ and should no be confused with the non-additivity parameter of
Tsallis thermostatistics, which is denoted here by $\bar{\tau}(q)\equiv
1/\tau(q)$.

Thermodynamic functions of the model under consideration can be found according
to the Tsallis non-extensive scheme \cite{Tsallis}. However, related
expressions are very cumbersome even in the simplest case of the ideal gas
\cite{3,4,5} and take the usual form only within the slightly non-extensive
limit \cite{6}. At the same time, developed scheme allows for to use
thermodynamic formalism of multifractal objects \cite{BS}, within the which the
multifractal exponent $q$ plays a role of a state parameter. If the dependence
$\tau(q)$ has some singularities, then variation in $q$ may arrive at phase
transitions. It is worthwhile to stress that developed scheme arrives directly
at related singularities of thermodynamic functions type of the internal energy
(see below Eq.(\ref{18})), the entropy (cf. Eq.(\ref{4}))
\begin{equation}
S_q={\bar\tau}(q)\ln_{2-{\bar\tau}(q)}\left(W_q\right),\quad{\bar\tau}(q)\equiv
1/\tau(q)\label{4aa}
\end{equation}
and the free energy
\begin{equation}
F_q=E_q-TS_q.\label{4b}
\end{equation}

According to Ref.\cite{3}, the physical distribution is not the
input probability (\ref{14}), but the escort one
\begin{equation}
\mathcal{P}(\varepsilon_i)\equiv\frac{P^{{\bar\tau}(q)}(\varepsilon_i)}
{\sum_{i=1}^{W_q}P^{{\bar\tau}(q)}(\varepsilon_i)}. \label{a}
\end{equation}
It corresponds to the condition
\begin{equation}
\sum\limits_{i=1}^{W_q}(\varepsilon_i-E_q)P^{{\bar\tau}(q)}(\varepsilon_i)=0
\label{b}
\end{equation}
instead of the second equation (\ref{13}). In difference of the distribution
(\ref{14}) related probability
\begin{equation}
\mathcal{P}(\varepsilon_i)=\mathcal{Z}_q^{-1}\exp_{\tau(q)}\left[-{\bar\tau}(q)
\left(\varepsilon_i-E_q\right)/T\right] \label{c}
\end{equation}
is determined with the physical temperature $T$.

In the case of continuous energy spectrum characterized with the density
distribution $\rho(\varepsilon)$, the internal energy related to the condition
(\ref{b}) takes the form
\begin{equation}
E_q=\int\limits_{-\infty}^{\infty}\varepsilon
\mathcal{P}(\varepsilon)\rho(\varepsilon){\rm d}\varepsilon. \label{16}
\end{equation}
Extreme value of $E_q$ is reached at the condition
\begin{equation}
\frac{\rho^{'}(\varepsilon)} {\rho(\varepsilon)}\simeq-\frac{
\mathcal{P}^{'}(\varepsilon)}{\mathcal{P}(\varepsilon)}
\label{17}
\end{equation}
where prime denotes differentiation over $\varepsilon$. Usually, the density
function is reduced to the power law $\rho(\varepsilon)\sim\varepsilon^{cN}$,
$c\sim 1$, so that $\rho^{'}(\varepsilon)/\rho(\varepsilon)\simeq
cN/\varepsilon$. Then, with using the distribution (\ref{c}), the condition
(\ref{17}) taken at $\varepsilon=E_q$ arrives at the equipartition law
\begin{equation}
E_q=c\tau(q)NT\label{18}
\end{equation}
where the value $c\tau(q)$ is the specific heat.

\section{Optimization of multifractal spectrum}\label{sec:level3}

\begin{figure}[!h]
\centering
 a\hspace{6cm}b\\
\includegraphics[width=60mm]{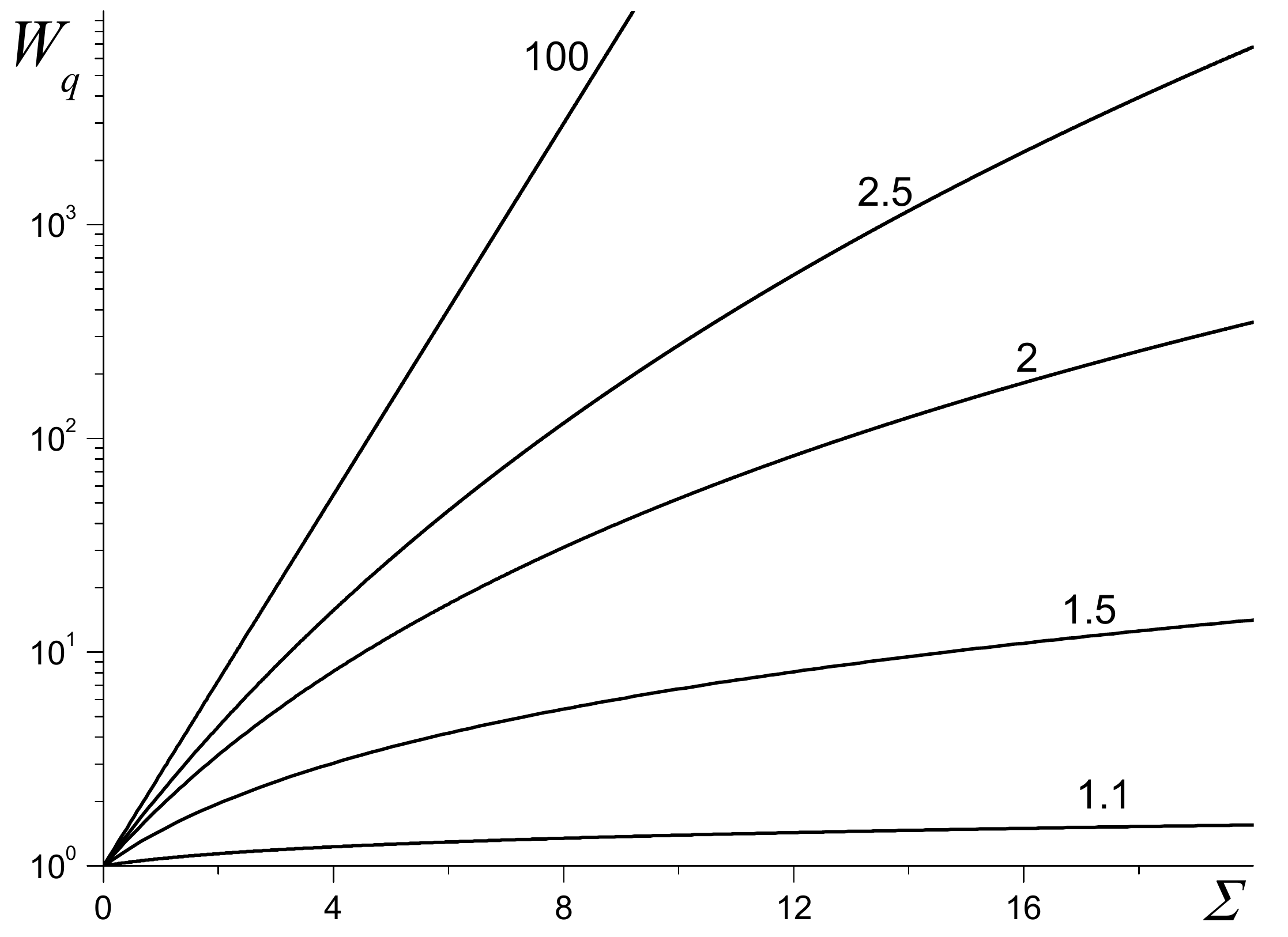}
\includegraphics[width=60mm]{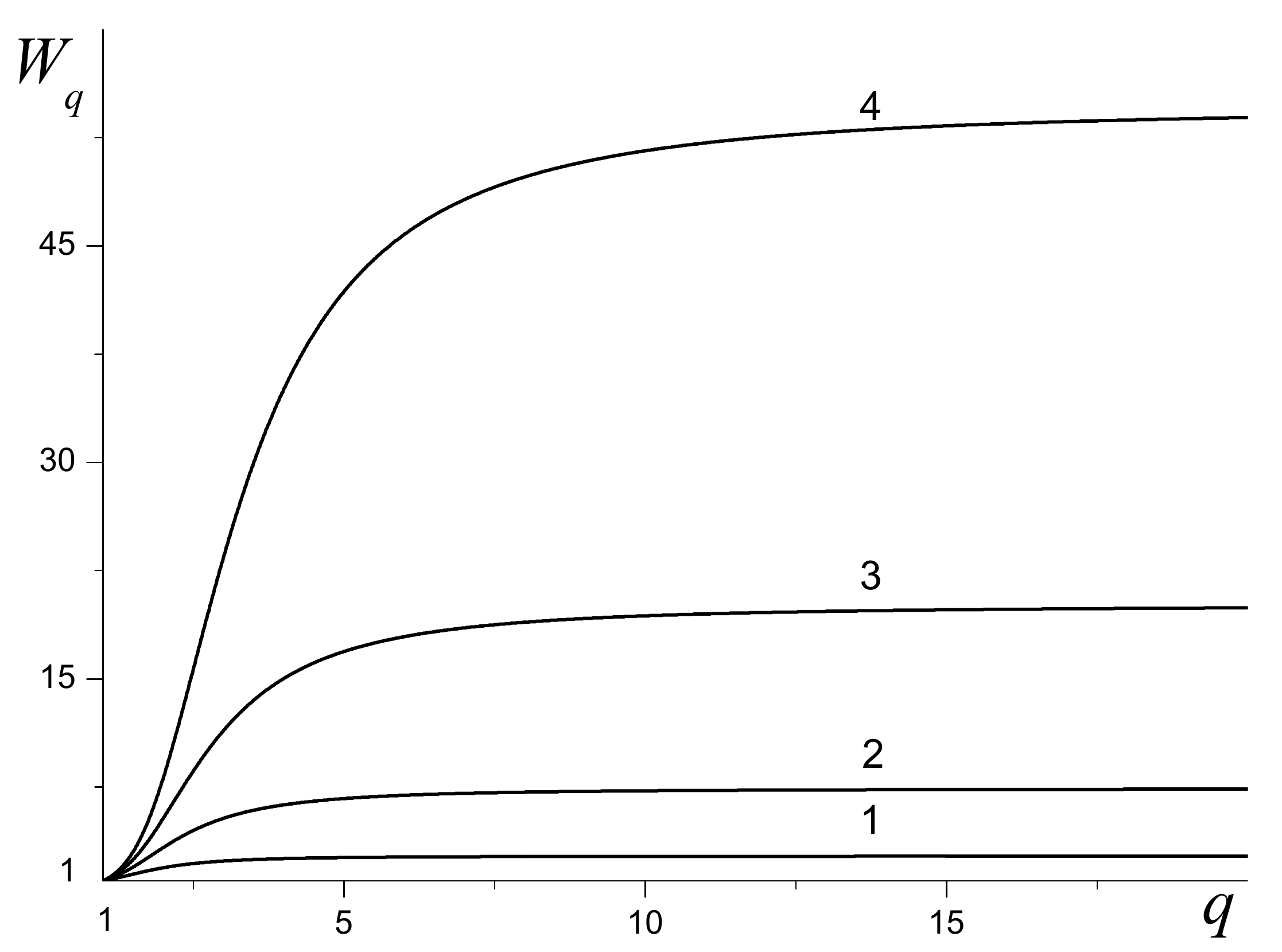}\\
\caption{Dependencies of the statistical weight on the complexity (a) and
multifractal exponent (b) at the exponent (\ref{23ab}) (numbers near
curves on the left panel show values $q$, on the right one -- values $\Sigma$)}
 \label{W_q}
\end{figure}

Up to now, we suppose that the multifractal spectrum $f(d)$ is arbitrary. If it
is optimized at normalization condition
\begin{equation}
\int\limits_0^1 f(d){\rm d}d=1, \label{19}
\end{equation}
one has to minimize the expression
\begin{equation}
\tilde{S}_q\{f(d)\}=\int\limits^{\gamma(W_q)}_{\gamma(1)}\left[\int\limits^1_0\gamma^{qd-f(d)}\rho(d){\rm
d}d\right]^{-1}{\rm d}\gamma-\frac{\Sigma ^2}{2}\left[\int\limits_0^1 f(d){\rm
d}d-1\right] \label{20}
\end{equation}
where we take into account Eqs. (\ref{2}), (\ref{3a}), and $\Sigma $ determines
the Lagrange multiplier. As a result, we arrive at the equality
\begin{equation}
\int\limits^{\gamma(W_q)}_{\gamma(1)}\gamma^{\left\{\left[qd-f(d)\right]-2\tau(q)\right\}}\ln\gamma~{\rm
d}\gamma=\frac{\Sigma ^2}{2}\label{21}
\end{equation}
whose integration gives, with accounting Eq.(\ref{4a}), the transcendental
equation
\begin{eqnarray}
\frac{1}{2}\left[\Sigma\tau(q){\mathcal T}_q(d)\right]^2-W_q^{{\mathcal T}
_q(d)}\left[{\mathcal T}_q(d)\ln(W_q)-1\right]-1=0,\label{22}\\ {\mathcal
T}_q(d)\equiv \bar{\tau}(q)\left\{1-2\tau(q)+\left[qd-f(d)\right]\right\}.
\label{23}
\end{eqnarray}
This equation is written in the form, when can be used either the spectrum
function $f(d)$ or the exponent dependence $\tau(q)$. In the latter case, we
find initially the dependence $q(d)$ from the equation
\begin{equation}
\left.\frac{{\rm d}\tau}{{\rm d}q}\right|_{q=q(d)}=d,
\label{10a}
\end{equation}
being conjugated to Eq.(\ref{6a}). Then, inserting this dependence into
Eq.(\ref{23}), we arrive at the trivial expression
\begin{equation}
{\mathcal T}(q)\equiv{\mathcal T}_q(d(q))=\bar{\tau}(q)-1\label{23a}
\end{equation}
whose using in Eq.(\ref{22}) allows for to determine the dependence of the
statistical weight $W_q$ versus the complexity $\Sigma$ at given function
$\tau(q)$. A typical form of this dependence at the mass exponent
\begin{equation}
\tau=\frac{q-1}{\sqrt{1+(q-1)^2}}
 \label{23ab}
\end{equation}
is shown in Fig.\ref{W_q}.
It is seen the statistical weight increases monotonically as with growth of the
complexity, so with increasing multifractal exponent.

In the limit of smooth phase space, when $q\to\infty$, $d\to 1$, ${\mathcal
T}(q)\to 0$, one obtains the usual expression for the statistical weight of the
complex system
\begin{equation}
W_\infty={\rm e}^{\Sigma_\infty},\quad \Sigma_\infty\equiv\sigma_\infty
N\label{24}
\end{equation}
which is determined by the specific complexity $\sigma_\infty$ per one
particle. At small deviation off the minimum complexity $(\Sigma
-\Sigma_\infty\ll\Sigma_\infty)$ and light multifractality
($1-\tau(q)\ll\Sigma_\infty^{-1}$), linearized equation (\ref{22}) gives
\begin{equation}\label{25}
\begin{split}
&W_q\simeq
W_{\tau(q)}\left\{1+\tau(q)\left[1-\frac{2}{3}\left(1-\tau(q)\right)\Sigma_\infty
\right]\left(\Sigma -\Sigma_\infty\right)\right\},\\
&W_{\tau(q)}\equiv\exp\left\{\tau(q)\Sigma_\infty\left[1-\frac{1}{3}\left(1-\tau(q)\right)
\Sigma_\infty\right]\right\},\quad 1\ll q<\infty.
\end{split}
\end{equation}
In the opposite case $\tau(q)\ll 1$, one has within logarithmic accuracy
\begin{equation}
W_q\simeq\left[\alpha\tau(q)\Sigma^2\right]^{\tau(q)}, \ \alpha\sim 1,\quad
q-1\ll 1. \label{24a}
\end{equation}

As show above findings, optimization of the multifractal spectrum, obeying the
normalization condition (\ref{19}), gives the dependence of the statistical
weight $W_q$ versus the system complexity $\Sigma$ at given multifractal
exponent $\tau(q)$. Naturally, relations (\ref{22}) and (\ref{23a}) allows for
to solve the inverse problem -- to find the dependence $\tau(q)$ at given
function $W_q(\Sigma)$. However, definition of the dependence $W_q(\Sigma)$
leads to very complicated problem. It is more convenient to use a modeling
function $\tau(q)$ bounded with asymptotics (\ref{11}) and then to find the
statistical weight $W_q$. Within this algorithm, in the following Section we
model the multifractal spectrum on the basis of the procedure of both Tsallis
and Kaniadakis deformations. It is appeared, such deformations give the whole
set of functions $f(d)$ to present all possible types of multifractal spectra.

\section{Analytical modeling multifractal spectrum}\label{sec:level4}

\begin{figure}[!h]
\centering
\includegraphics[width=80mm]{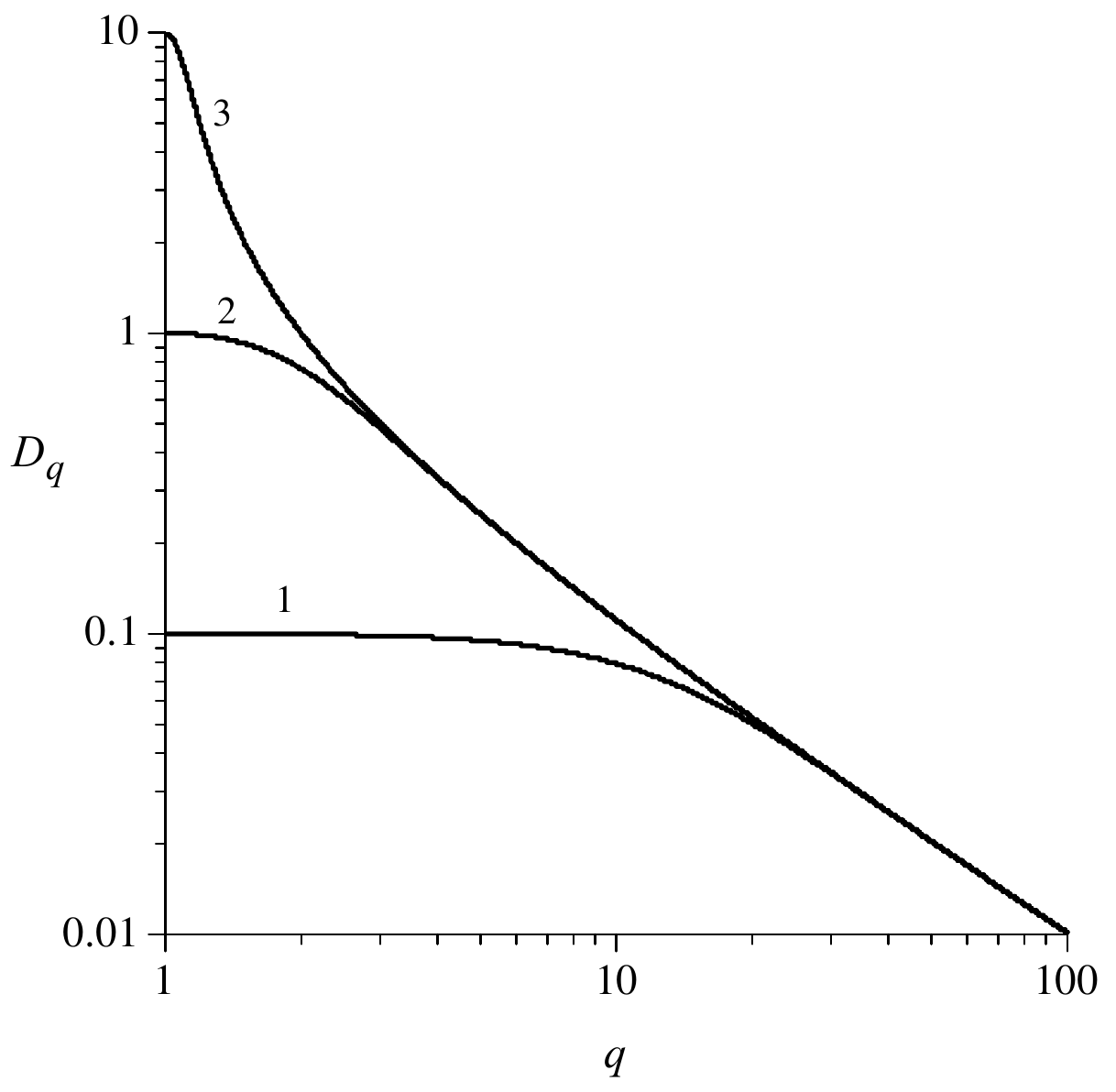}
\caption{Multifractal
dimension function related to the dependence (\ref{a8}) (curves 1,
2, 3 correspond to $\mathcal D=0.1, 1.0, 10$).}\label{Dq(q)_D}
\end{figure}

In the simplest case, we can take the function $\tau(q)$ in the form
\begin{equation}
\tau=\tanh\left[\mathcal D(q-1)\right]\label{a8}
\end{equation}
being determined by parameter $\mathcal D>0$ and argument
$q\in[1,\infty]$. According to Fig.\ref{Dq(q)_D} related
multifractal dimension function
\cite{Feder}
\begin{equation}
D_q=\frac{\tau(q)}{q-1} \label{a9}
\end{equation}
monotonically decreases from the maximum value $D_0=\mathcal D$ to the minimum
one $D_\infty=0$ with increase in $q$. However, the maximum value of the
fractal dimension $D_q$ is fixed by the magnitude $D_0=1$, so that one should
put $\mathcal D=1$ in the dependence (\ref{a8}). As a result, it takes quite
trivial form.

Due to the $\tau(q)$ function increases monotonically within narrow interval
$[0,1]$, one has a scanty choice of its analytical models. To set a possible
representation of $\tau(q)$ one can use a deformation of the hyperbolic tangent
(\ref{a8}) at $\mathcal D=1$. By now, two analytical procedures of such
deformation are extensively popularized. The first of them is based on the
Tsallis exponential form \cite{Tsallis}
\begin{eqnarray}
\exp_{\kappa}(x)\equiv\left\{
\begin{array}{ll}
\left(1+\kappa x\right)^{1/\kappa}\quad\ {\rm at}\quad 1+\kappa x>0,\\ 0\ \
\qquad\quad\qquad\qquad\quad{\rm otherwise}
\end{array} \right.
\label{a10}
\end{eqnarray}
where deformation parameter $\kappa$ takes positive values. The second
procedure has been proposed by Kaniadakis \cite{Kaniadakis} to determine the
deformed exponential form
\begin{equation}
\exp_{\kappa}(x)\equiv\left(\kappa x+\sqrt{1+\kappa^2 x^2}\right)^{1/\kappa}.
\label{a11}
\end{equation}
%With
With using these definitions, the deformed tangent (\ref{a8}) takes the form
\begin{equation}
\tau_{\kappa}(q)=\tanh_{\kappa}(q-1)\equiv\frac{\exp_{\kappa}(q-1)-
\exp_{\kappa}(1-q)}{\exp_{\kappa}(q-1)+\exp_{\kappa}(1-q)}\label{a12}
\end{equation}
where the multifractal exponent $q$ varies within the domain $[1,\infty]$.

\begin{figure}[!h]
\centering
\includegraphics[width=80mm]{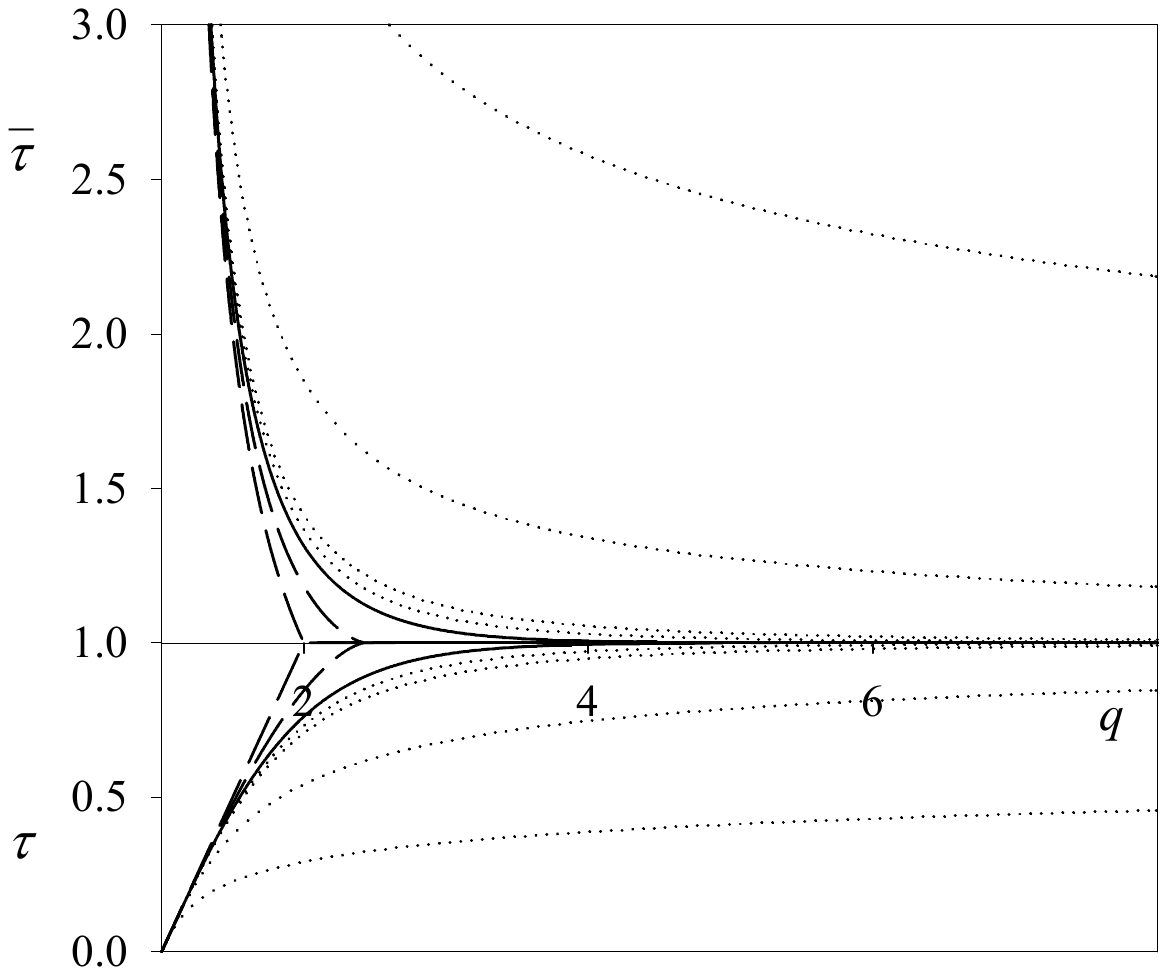}
\caption{The $q$-dependencies of the multifractal exponent $\tau$ and its
inverse value $\bar\tau=1/\tau$ (solid line corresponds to $\kappa=0$, dashed
curves relate to the Tsallis deformation with $\kappa=0.7, 1$; dotted lines
correspond to the Kaniadakis one at $\kappa=0.7, 1, 3, 10$).}\label{tau(q)}
\end{figure}

\begin{figure}[!h]
\centering
\includegraphics[width=60mm]{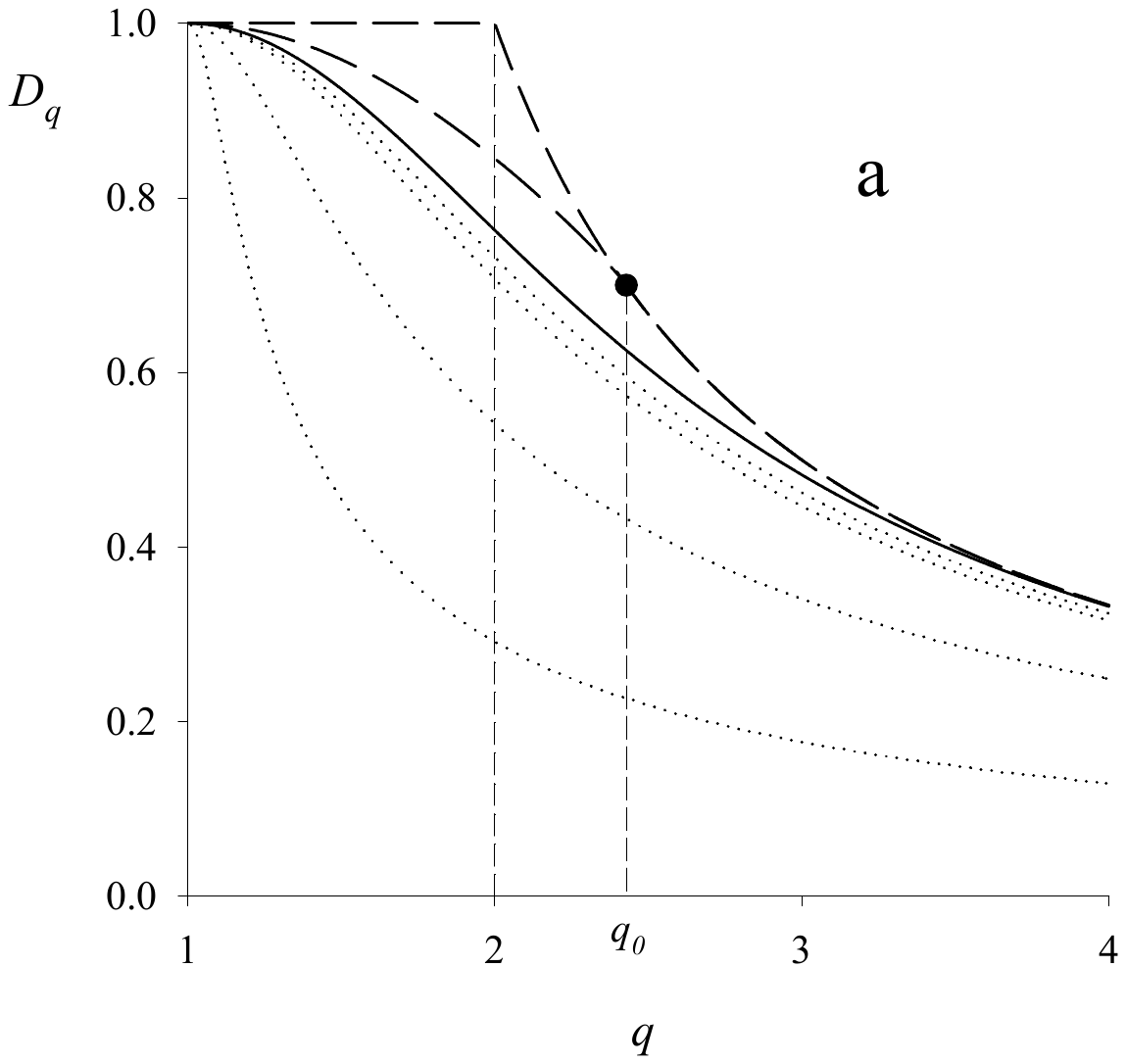}
\hspace{0.5cm}\includegraphics[width=60mm]{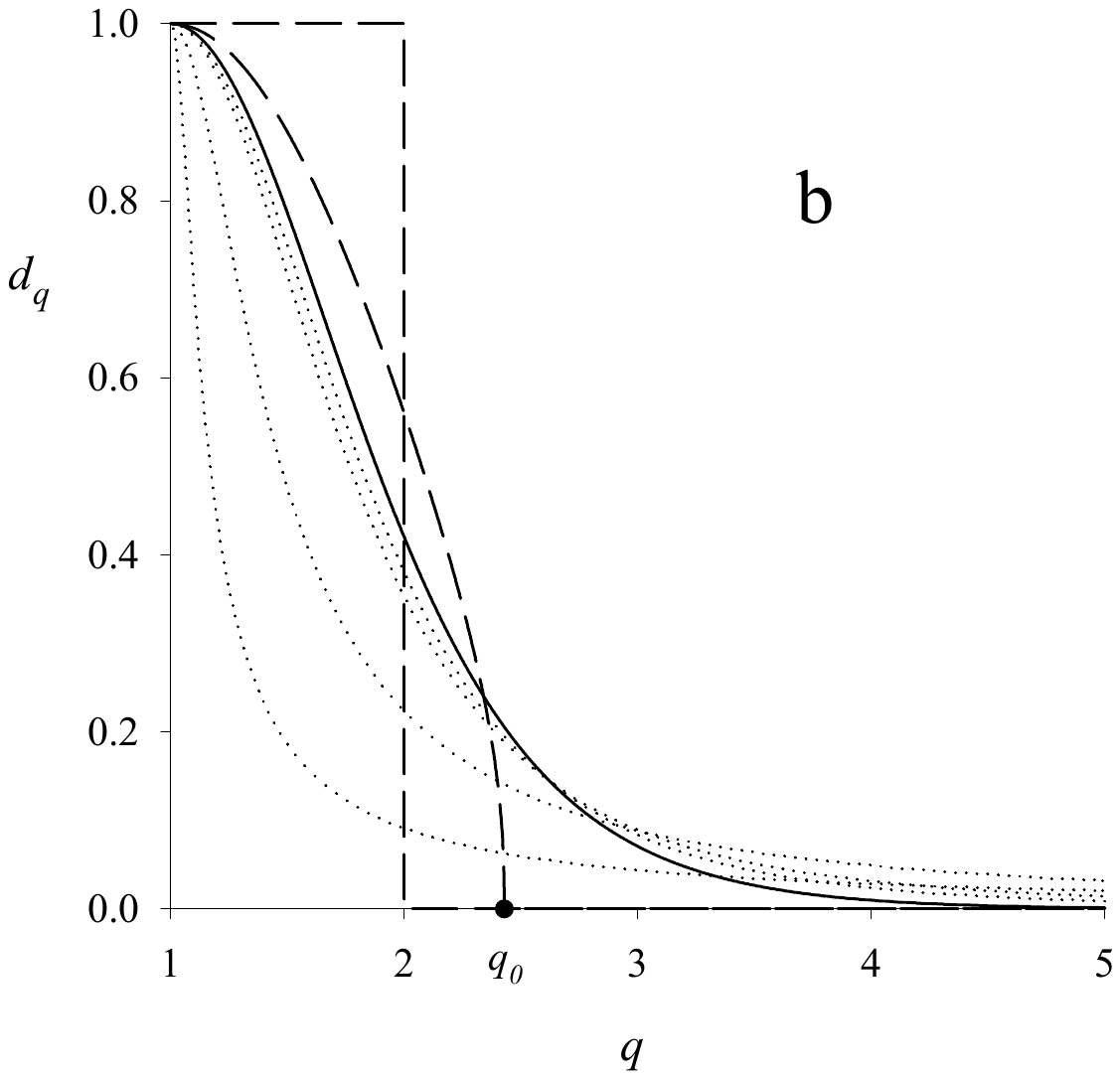} \caption{Spectra
of fractal (a) and specific (b) dimensions of the phase space (solid
line corresponds to $\kappa=0$, dashed curves relate to the Tsallis
deformation with $\kappa=0.7, 1$; dotted lines correspond to the
Kaniadakis one at $\kappa=0.7, 1, 3, 10$).} \label{d_q}
\end{figure}

The $q$-dependencies of the multifractal exponent $\tau(q)$ and its inverse
value $\bar\tau(q)=1/\tau(q)$ are shown in Fig.\ref{tau(q)} at different
magnitudes
of both Tsallis and Kaniadakis deformation parameters $\kappa$. (The $\kappa$
values are picked out in such a manner to cover uniformly the panels of Figs.
\ref{tau(q)}--\ref{f(d)} with related curves.) It is principally important, the
first of these deformations arrives at more fast variations of both exponents
$\tau(q)$ and $\bar\tau(q)$ in comparison with non-deformed hyperbolic tangent
$\tau_0=\tanh_0(q-1)$, whereas the Kaniadakis deformation slows down these
variations with $\kappa$ increase.

A characteristic peculiarity of the Tsallis deformation consists in breaking
dependencies $\tau(q)$, $\bar\tau(q)$ in the point $q_0=(1+\kappa)/\kappa$
where the second terms in both numerator and denominator of the definition
(\ref{a12}) take the zero value. As a result, the multifractal exponent
(\ref{5a}) takes the asymptotics
\begin{eqnarray}
\tau_{\kappa}^{(Ts)}\simeq\left\{
\begin{array}{ll}
\left(q-1\right)-\frac{1-\kappa^2}{3}\left(q-1\right)^3 \quad\qquad {\rm
at}\quad\ 0<q-1\ll 1,\\
1-2\left(\kappa/2\right)^{1/\kappa}\left(q_0-q\right)^{1/\kappa}\quad\quad {\rm
at}\quad 0<q_0-q\ll q_0.
\end{array} \right.
\label{a13}
\end{eqnarray}
For $\kappa=1$ the dependence $\tau_1^{(Ts)}(q)$ takes the simplest form:
$\tau_1^{(Ts)}=q-1$ at $1\leq q\leq 2$, and $\tau_1^{(Ts)}=1$ at $q>2$. It is
worthwhile to note, the Tsallis deformation parameter can not take values
$\kappa>1$ because these relate to the fractal dimensions $D_q>1$ at $q\ne 0$.

In the case of the Kaniadakis deformation, the multifractal exponent
$\tau_{\kappa}(q)$ varies smoothly to be characterizes by the following
asymptotics:
\begin{eqnarray}
\tau_{\kappa}^{(K)}\simeq\left\{
\begin{array}{ll}
\left(q-1\right)-\frac{2+\kappa^2}{6}\left(q-1\right)^3 \qquad {\rm at}\quad \
0<q-1\ll 1,\\ 1-2\left[2\kappa(q-1)\right]^{-2/\kappa}\quad\quad\quad {\rm
at}\quad\quad \kappa(q-1)\gg 1.
\end{array} \right.
\label{a14}
\end{eqnarray}
In contrast to the Tsallis case, here the deformation parameter can take
arbitrary values to give the simplest dependence (\ref{23ab}) at $\kappa=1$.

The fractal dimension (\ref{a9}) as a function of the $q$ exponent falls down
monotonically as shown in Fig.\ref{d_q}a. According to Eqs.(\ref{a13}), in the
case of the
Tsallis deformation, one has a broken dependence $D(q)$, being characterized by
the asymptotics
\begin{eqnarray}
D_q^{(Ts)}\simeq\left\{
\begin{array}{ll}
1-\frac{1-\kappa^2}{3}\left(q-1\right)^2 \quad\quad\quad\qquad {\rm at}\quad\ \
0<q-1\ll 1,\\
\frac{1}{q-1}-2\left(\kappa/2\right)^{1/\kappa}\frac{\left(q_0-q\right)^{1/\kappa}}
{q-1}\quad\quad {\rm at}\quad 0<q_0-q\ll q_0.
\end{array} \right.
\label{a15}
\end{eqnarray}
In the limit case $\kappa=1$, the phase space is smooth
$(D_q^{(Ts)}=1)$ within the interval $1\leq q \leq 2$. For the
Kaniadakis deformation, the fractal dimension $D_q^{(K)}$ is given
by smoothly falling down curve whose slope increases with the
deformation parameter growth. According to Eqs.(\ref{a14}), in this
case, one has the asymptotics
\begin{eqnarray}
D_q^{(K)}\simeq\left\{
\begin{array}{ll}
1-\frac{2+\kappa^2}{6}\left(q-1\right)^2 \quad\quad\qquad\qquad\qquad {\rm
at}\quad \ 0<q-1\ll 1,\\
(q-1)^{-1}-2\left(2\kappa\right)^{-2/\kappa}(q-1)^{-(\kappa+2)/\kappa}\ \ {\rm
at}\ \ \kappa(q-1)\gg 1.
\end{array} \right.
\label{a16}
\end{eqnarray}
At $\kappa=1$, the typical dependence related to Eq.(\ref{23ab}) is of the form
\begin{equation}
D_q^{(K)}=\left[1+(q-1)^2\right]^{-1/2}. \label{a17c}
\end{equation}

\section{Multifractal spectrum}\label{sec:level5}

At given multifractal exponent $\tau(q)$, the spectrum function $f(d)$ is
defined by the Legendre transformation (\ref{5a}) where the specific
dimension reads
\begin{equation}
d_q=\frac{{\rm d}\tau}{{\rm d}q}. \label{a17}
\end{equation}
As shows Fig.\ref{d_q}b, the dependence $d_q$ has monotonically falling down
form to take the value $d_q=0$ at $q>q_0\equiv(1+\kappa)/\kappa$ for the
Tsallis deformation. In this case, asymptotical behaviour is characterized by
Eqs.(\ref{a13}), according to which one obtains
\begin{eqnarray}
d_q^{(Ts)}\simeq\left\{
\begin{array}{ll}
1-(1-\kappa^2)\left(q-1\right)^2 \quad\quad\quad\quad\ \ {\rm at}\quad\
0<q-1\ll 1,\\
\left(\kappa/2\right)^{(1-\kappa)/\kappa}\left(q_0-q\right)^{(1-\kappa)/\kappa}\qquad
{\rm at}\quad 0<q_0-q\ll q_0.
\end{array} \right.
\label{a18}
\end{eqnarray}
In the limit $\kappa\to 1$, the dependence $d^{(Ts)}(q)$ takes the step-like
form being $d_q=1$ within the interval $1\leq q\leq 2$ and $d_q=0$ otherwise.

For the Kaniadakis deformation, Eqs.(\ref{a14}) arrive at the asymptotics
\begin{eqnarray}
d_q^{(K)}\simeq\left\{
\begin{array}{ll}
1-\frac{2+\kappa^2}{2}\left(q-1\right)^2 \quad\quad\qquad\qquad {\rm at}\quad \
0<q-1\ll 1,\\
2^{2(\kappa-1)/\kappa}\left[\kappa(q-1)\right]^{-(2+\kappa)/\kappa}\quad\quad
{\rm at}\quad\quad \kappa(q-1)\gg 1.
\end{array} \right.
\label{a19}
\end{eqnarray}
The typical behaviour is presented by the dependence
\begin{equation}
d_q^{(K)}=\left[1+(q-1)^2\right]^{-3/2} \label{a5ab}
\end{equation}
related to $\kappa=1$.

The multifractal spectrum is defined by the equality
\begin{equation}
f(d)=dq_d-\tau(q_d)\label{a5a}
\end{equation}
being conjugated to Eq.(\ref{5a}). Here, the specific multifractal exponent
$q_d$ is determined by Eq.(\ref{a17}) which arrives at the limit relations
(\ref{a18}), (\ref{a19}). With their using, one obtains the asymptotics
\begin{eqnarray}
f^{(Ts)}\simeq\left\{
\begin{array}{ll}
d-\frac{2}{3}(1-\kappa^2)^{-1/2}(1-d)^{3/2} \quad\quad\quad\ \ {\rm at}\quad\
0<1-d\ll 1,\\ -\left[1+2(1/\kappa-1)d^{1/(1-\kappa)}\right]+(1+1/\kappa)d\qquad
{\rm at}\quad\ d\ll 1
\end{array}
\right. \label{a20}
\end{eqnarray}
for the Tsallis deformation, and the relations
\begin{eqnarray}
f^{(K)}\simeq\left\{
\begin{array}{ll}
d-\frac{2}{3}\left(1+\frac{\kappa^2}{2}\right)^{-1/2}(1-d)^{3/2} \quad\quad\ \
{\rm at}\quad 0<1-d\ll 1,\\
-\left[1-2^{(\kappa-4)/(2+\kappa)}(1+2/\kappa)d^{2/(2+\kappa)}\right]+d\quad\
{\rm at}\quad d\ll 1,
\end{array} \right.
\label{a21}
\end{eqnarray}
characterizing the Kaniadakis deformation.

\begin{figure}[!h]
\centering
\includegraphics[width=80mm]{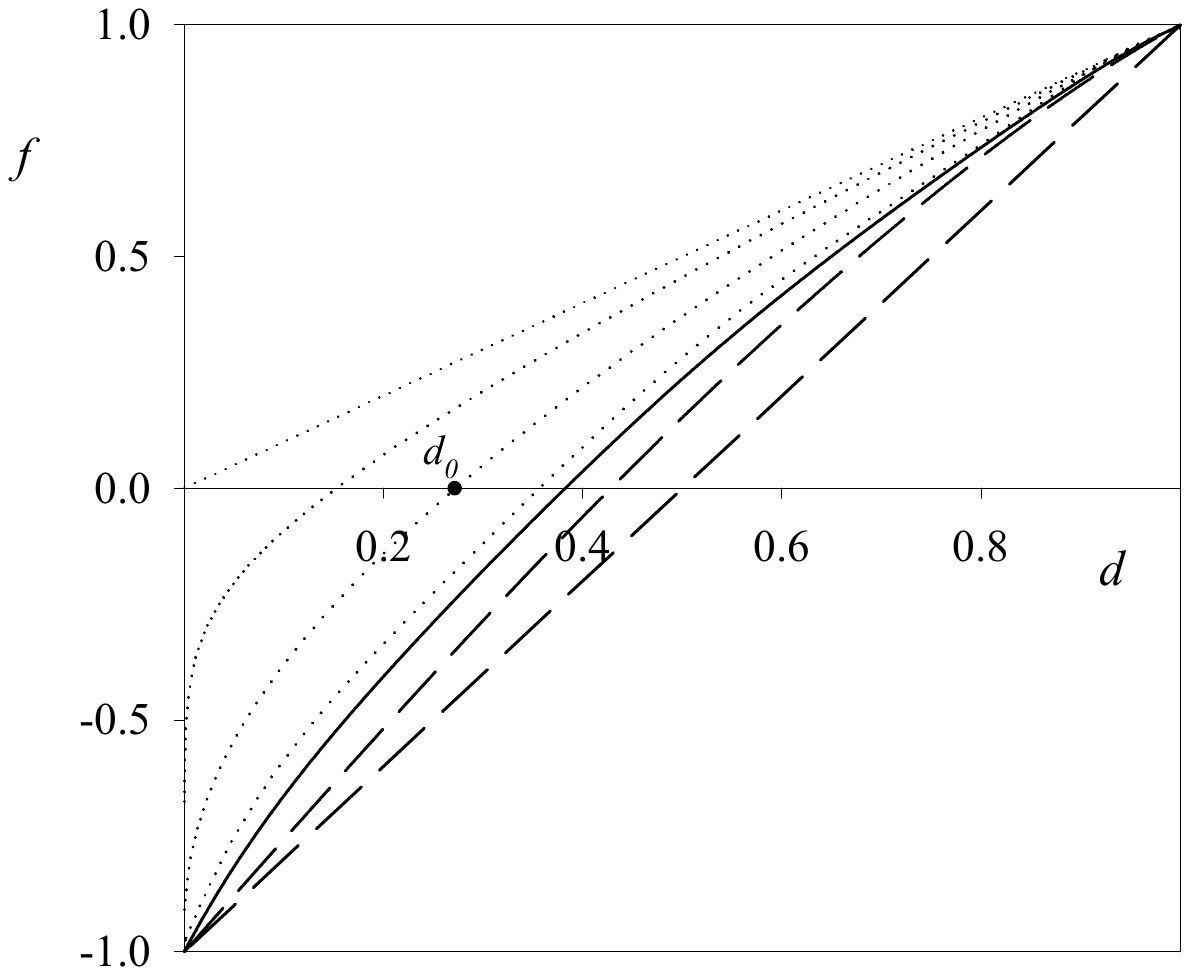}
\caption{Spectrum function of the multifractal phase space (solid line
corresponds to $\kappa=0$, dashed curves relate to the Tsallis deformation with
$\kappa=0.7, 1$; dotted lines correspond to the Kaniadakis one at $\kappa=1, 3,
10,\infty$).}\label{f(d)}
\end{figure}

As shows Fig.\ref{f(d)}, for finite deformation parameters $\kappa<\infty$,
a spectrum function increases monotonically, taking the minimum value $f=-1$ at
$d=0$ and the maximum one $f=1$ at $d=1$. Besides, the derivative $f'\equiv{\rm
d}f/{\rm d}d$ equals to $f'(0)=\infty$ on the left boundary and $f'(1)=1$ on
the right one. It is significant, the whole set of the spectrum functions is
bounded by the limit dependencies $f^{(Ts)}=2d-1$ and $f^{(K)}=d$, the first of
which relates to limit magnitude of the Tsallis deformation parameter
$\kappa=1$ whereas the second one corresponds to the Kaniadakis limit
$\kappa=\infty$. A typical form of the spectrum function is presented by the
dependencies
\begin{eqnarray}
f^{(K)}=\left\{
\begin{array}{ll}
-d\ln\left(\frac{\sqrt{d}}{1+\sqrt{1-d}}\right)+\left(d-\sqrt{1-d}\right)\quad
{\rm at}\quad
\kappa=0,\\-\left(1-d^{2/3}\right)^{3/2}+d\quad\quad\quad\quad\quad\quad\ \
{\rm at}\quad\kappa=1.
\end{array} \right.
\label{a22}
\end{eqnarray}

\begin{figure}[!h]
\centering
\includegraphics[width=70mm]{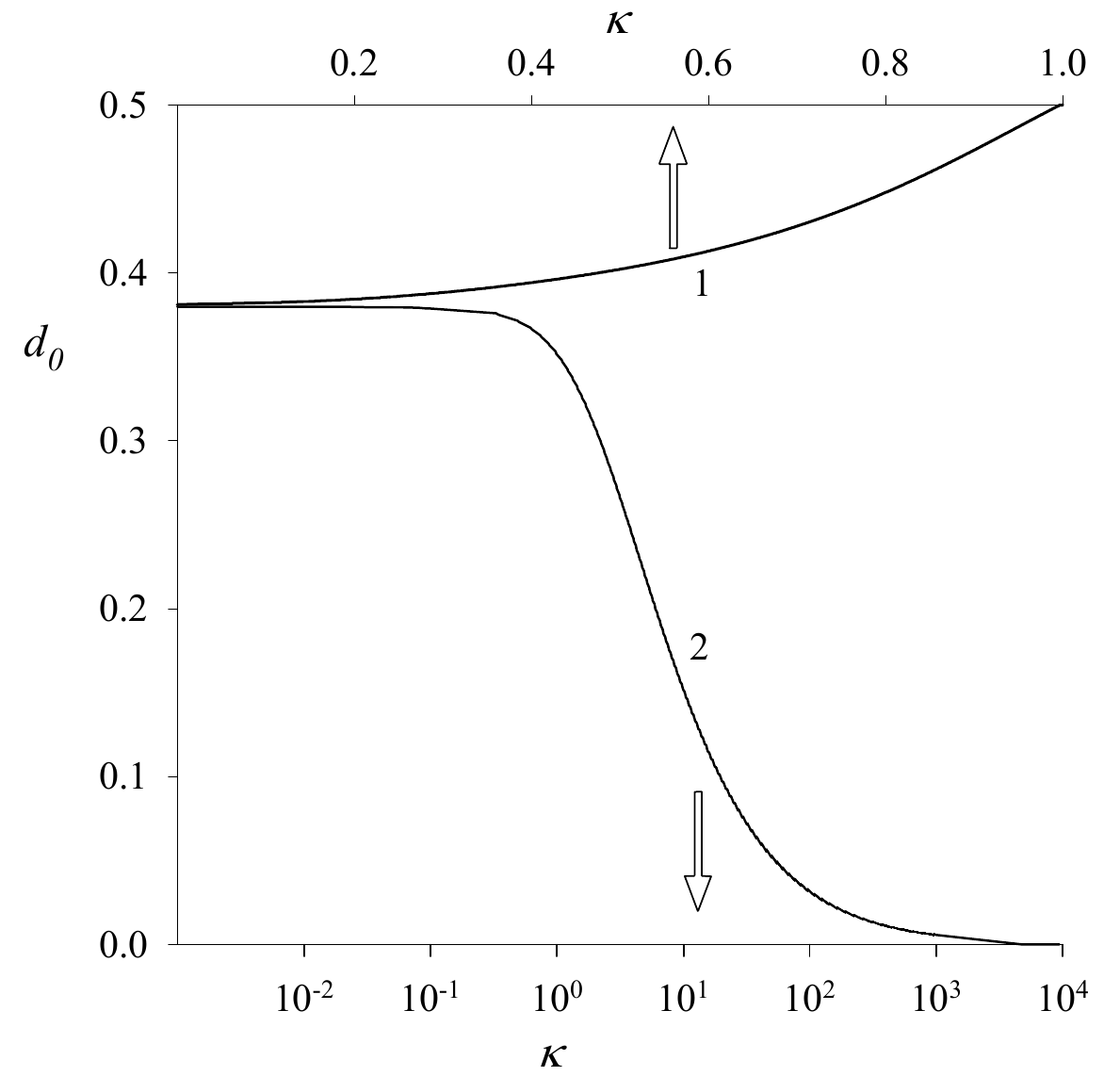}
\caption{Dimension $d_0$ related to the condition $f(d)=0$ as function of the
parameter $\kappa$ (curve 1 corresponds to the Tsallis deformation, curve 2 --
to the Kaniadakis one; positive values $f(d)$ relate to the domain
$d>d_0$).}\label{d0(k)}
\end{figure}

It may seem, at the first glance, that negative values of the spectrum function
$f(d)$ has not a physical meaning. To clear up this problem, let us take the
set of monofractals with the specific dimension $d=0$. Obviously, such
monofractals relate to the whole set of the phase space points, whose number
equals to the dimensionless volume $\gamma$. Just such result gives the
definition (\ref{2}) in the point $d=0$ where $f=-1$. On the other hand, in
opposite case $d=1$ where $f=1$, we obtain the very thing the number of
monofractals with volume $\gamma$ equals $\mathcal{N}_1=\gamma^{-1}$ to give
one multifractal of the same volume $\gamma$. At the same time, a single
monofractal is contained in the multifractal at condition $f(d)=0$ which takes
place at the specific dimension $d_0$ whose dependence on the deformation
parameter $\kappa$ is shown in Fig.\ref{d0(k)}.

\begin{figure}[!h]
\centering
\includegraphics[width=95mm]{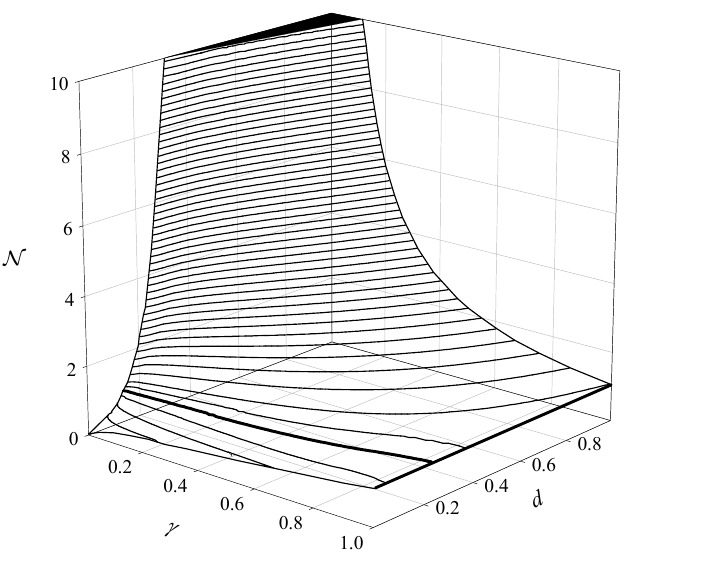}
\caption{The number $\mathcal{N}$ of monofractals within the phase space volume
$\gamma$ related to the specific dimension $d$ at the deformation parameter
$\kappa=10$ (different levels $\mathcal{N}={\rm const}$ are shown with thin
curves, thick lines relate to $\mathcal{N}=1$).}\label{N(k)}
\end{figure}

The dependence of the number $\mathcal{N}$ of monofractals containing in the
phase space volume $\gamma$ related to the multifractal with the specific
dimension $d$ is shown in Fig.\ref{N(k)}. It is seen, the number
$\mathcal{N}$ increases with the $\gamma$ volume growth at small dimensions
$d$, whereas in the limit $d\to 1$ the dependence $\mathcal{N}(\gamma)$ becomes
falling down to give infinitely increasing numbers $\mathcal{N}$ at $\gamma\to
0$. The speed of such increase growths monotonically with both decrease of the
Tsallis deformation parameter $\kappa$ and increase of the Kaniadakis one.

\section{Conclusions}\label{sec:level6}

As above consideration shows, the statistical mechanics of self-similar complex
systems with phase space, whose specific fractal dimension $d$ is distributed
with spectrum $f(d)$, is governed by the Tsallis formalism of the non-extensive
thermostatistics. In this way, the role of non-additivity parameter plays
inverse value of the multifractal function $\tau(q)= qd_q-f(d_q)$ which
monotonically increases, taking value $\tau=0$ at $q=1$ and $\tau\simeq 1$ at
$q\to\infty$ (the latter limit relates to the smooth phase space, where
$\tau=1$). The multifractal function $\tau(q)$ is reduced to the specific heat
to determine, together with the inverse value ${\bar\tau}(q)\equiv
1/\tau(q)>1$, both statistical distributions and thermodynamic functions of the
system under consideration. At given function $\tau(q)$, optimization of the
normalized multifractal spectrum $f(d)$ arrives at the dependence of the
statistical weight on the system complexity.

It is shown the whole set of monofractals within a multifractal related to the
phase space, which gives the support of a generalized thermostatistics, is
modeled by the mass exponent $\tau(q)$ that determines the statistical weight
(\ref{4a}) at given volume $\gamma$. To be the entropy (\ref{2}) concave,
Lesche stable et cetera, the exponent $\tau(q)$ should be a function,
monotonically increasing within the interval $[0,1]$ at multifractal exponent
variation within the domain $[1,\infty]$. The simplest case of such a function
gives the hyperbolic tangent $\tau=\tanh(q-1)$ whose deformation (\ref{a12})
defined in accordance with both Tsallis and Kaniadakis exponentials
(\ref{a10}), (\ref{a11}) allows for to describe arbitrary multifractal phase
space explicitly. At the same time, the Tsallis deformation arrives at more
fast variations of the statistical weight exponent $\tau(q)$ in comparison with
non-deformed hyperbolic tangent, whereas the Kaniadakis one slows down these
variations with increasing the deformation parameter $\kappa$. All possible
dependencies $\tau(q)$ are bounded from above by the linear function
$\tau^{(Ts)}=q-1$ at $q\in [1,2]$ which is transformed into the constant
$\tau=1$ at $q>2$. This dependence relates to the smooth phase space within the
Tsallis interval $q\in [1,2]$.

The dependence (\ref{2a}) of the number of monofractals within the phase space
volume $\gamma$ related to the multifractal with the specific dimension $d$ is
determined by the spectrum function $f(d)$. This function increases
monotonically, taking the minimum value $f=-1$ at $d=0$ and the maximum one
$f=1$ at $d=1$; besides, its derivative equals $f'(0)=\infty$ on the left
boundary and $f'(1)=1$ on the right one. The whole set of the spectrum
functions is bounded by the limit dependencies $f^{(Ts)}=2d-1$ and $f^{(K)}=d$,
the first of which relates to limit magnitude of the Tsallis deformation
parameter $\kappa=1$ and the second one corresponds to the Kaniadakis limit
$\kappa=\infty$. The number of monofractals within the multifractal increases
with the $\gamma$ volume growth at small dimensions $d$ and falls down in the
limits $d\to 1$ to give infinitely increasing at $\gamma\to 0$.

%\section{Acknowledgements}\label{sec:level7}
%
%We thank anonymous referees for constructive criticism.

\end{document}